\documentclass{article}
\usepackage{graphicx}
\usepackage{amsmath,amssymb}
\usepackage{graphics,color,array,calc,rotating,epsfig,psfrag}
\numberwithin{equation}{section}
\usepackage{cite}
\usepackage{bm}
\usepackage{dcolumn}  
\usepackage{amsfonts}
\usepackage[mathscr]{eucal}
\def\be{\begin{equation}} \def\ee{\end{equation}}
\def\bea{\begin{eqnarray}} \def\eea{\end{eqnarray}}

\newcommand\prt{\partial}

\newcommand{\nn}{\nonumber}
\begin{document}
\baselineskip 18pt%
\begin{titlepage}
\vspace*{1mm}%
\hfill%
\vspace*{15mm}%
\hfill
\vbox{
    \halign{#\hfil         \cr
          } 
      }  
\vspace*{20mm}

\centerline{{\large {\bf Entropy product of rotating black holes in three-dimensions}}}
\vspace*{5mm}
\begin{center}
{Davood Mahdavian Yekta \footnote{d.mahdavian@hsu.ac.ir}}\\
\vspace*{0.2cm}
{ Department of Physics, Hakim Sabzevari University, P.O. Box 397, Sabzevar, Iran}\\
\vspace*{0.1cm}
\end{center}

\begin{abstract} 
It has been shown that the product of the entropies of the inner Cauchy and outer event horizon of the charged axisymmetric and stationary black holes is a universal formula, which is independent of the black hole's mass. In this paper, we investigate this universality for the two kinds of rotating black holes in the three-dimensional gravity models. In fact, we study the spacelike warped anti-de Sitter black hole in the new massive gravity and the Ba\~{n}ados, Teitelboim, and Zanelli black hole in the minimal massive gravity. We show that this rule is held in the first theory. By contrast, in the latter case which includes a holographic gravitational anomalous term, we obtain that the universality does not work and the product depends on the mass. As a complement to the above verification, we also study the thermodynamic properties of these black holes.  
\end{abstract} 

\end{titlepage}

\section{INTRODUCTION}
The black holes can be expected to play an essential role in understanding the quantum theory of gravity in any dimension, which their mechanics is governed by the four laws of thermodynamics\cite{Bardeen:1973gs} (known as the black hole thermodynamics). About 50 years ago, Bekenstein and Hawking asserted that the entropy of black holes at the microscopic level is proportional to the area of their event horizons 
\cite{Bekenstein:1972tm,Hawking:1974sw}. A qualitative demonstration of this assertion has been done in string theory by counting the microscopic degrees of freedom\cite{Strominger:1996sh}. In all of the theories, whether containing string theory or gravity, the results depend on the existence of an anti-de Sitter(AdS) factor in the background or in the near horizon of solutions\cite{Sen:2005wa},\cite{Kraus:2005vz}(see also Refs. in\cite{Kraus:2005vz}). 

Even though most of the physical quantities of black holes are related to their outer event horizons, recently, the inner Cauchy horizon mechanics are also important\cite{Ansorg:2008bv,Ansorg:2009yi}. It seems that studying the properties of the inner horizon may help us better understand the nonextremal charged and rotating solutions in string theory and gravity\cite{Cvetic:1996kv}-\cite{Bento:1995qc}. There are similar intensive thermodynamical quantities for the inner Cauchy horizon as outer event horizon, which can illuminate whether the first law of black hole thermodynamics works for them or not\cite{Cvetic:2010mn}-\cite{Castro:2012av}. 

It has been shown in Refs.\cite{Ansorg:2009yi} and\cite{Cvetic:2010mn} that we can construct a quantity from the entropies of nonextremal black holes, which is independent of the continuous parameters in the theory, such as the black hole's ADM mass. This is the entropy product of black holes, which is called the entropy-product law (EPL) or inner mechanics. In other words, the product of the entropies for the inner and outer horizons of charged rotating black holes, i.e. $S^{+}\,S^{-}$, is independent of the black hole's mass and only depends on the quantized parameters, such as angular momentums or electromagnetic charges. However, on the conformal field theory (CFT) side, this corresponds to the level matching condition of the left- and right-moving modes such as $(N_{L}\,-\,N_{R})$ \cite{Larsen:1997ge}-\cite{Cvetic:1997xv}. Most of the studies have been done on the EPL in four- and five-dimensional theories, with some of them are in string theory as charged solutions of supergravity and heterotic theories\cite{Cvetic:1996kv},\cite{Cvetic:2005zi}, and others are in asymptotically flat solutions of pure gravity such as Kerr-Newman, rotating charged black rings, and black strings\cite{Castro:2012av}. 

In this paper, we investigate the EPL for two kinds of higher-derivative gravitational models in three dimensions.
The reason to study the higher-derivative theories of gravity is that they have been attracting much attention in recent years in the context of quantum gravity. It is known that such higher-derivative theories in four dimensions typically suffer from the ghost instability and unitarity problems\cite{Stelle:1976gc},\cite{Stelle:1977ry}. However, we are interested in three-dimensional gravity theories to better understand the implications of quantum theory of gravity. As another motivation, since pure gravity in three dimensions has no physical degrees of freedom, we pursue the higher-derivative gravities or cosmological models. 

In this sense, new massive gravity(NMG)\cite{Bergshoeff:2009hq} may be considered as the simplest higher curvature theory of gravity and expected to possess some essential properties, which could be inherited to the higher dimensional Gauss-Bonnet counterpart.
Though NMG is a ghost-free\cite{Bergshoeff:2009tb,Bergshoeff:2009aq} and  unitary\cite{Nakasone:2009bn} toy model around the flat background, it was shown that the theory suffers from
ghosts around the AdS background except for
the case with a fine-tuning between the coupling constants\cite{Liu:2009bk}. As the second theory, minimal massive gravity(MMG)\cite{Bergshoeff:2014pca} is a four-derivatives theory at the level of equations of motion, which includes one massive degree
of freedom similar to the topologically massive gravity(TMG)\cite{Deser:1982vy} theory. Not only this is a unitary theory in the bulk, but also it has a unitary dual CFT on the boundary for some values in the parameter space of the theory\cite{Bergshoeff:2014pca}.
A particular feature of this theory is that it has no
Lagrangian in the metric formalism, while is formulated in the first-order canonical structure\cite{Bergshoeff:2014pca},\cite{Yekta:2015gja}.

 The most important feature of these three-dimensional models is that they have black hole solutions similar to their higher dimensional counterparts in string theory or gravity. In fact, we will focus on the spacelike warped AdS (WAdS$_3$) black hole\cite{Moussa:2003fc}-\cite{Clement:2009gq} in the NMG theory and the Ba\~{n}ados, Teitelboim, and Zanelli(BTZ) black hole\cite{Banados:1992wn} in the MMG theory. The analysis for both solutions in TMG was discussed in Ref.\cite{Detournay:2012ug}. In the first case, we will show that the EPL is proportional to the angular momentum of the WAdS$_3$ black hole, whereas in the latter, this does not happen for the BTZ black hole. This behavior is referred to the diffeomorphism anomaly\cite{Kraus:2005zm} in the MMG theory. We also discuss about the thermodynamical quantities of the inner and event horizons in each sector and show that they satisfy the first law of black hole thermodynamics and Smarr-like formula for the mass.

The organization of the paper is as follows; In Sec. 2, we first introduce the NMG theory and then write the WAdS$_3$ black hole as a solution of the equations of motion for this theory. In the other subsections, we find the conserved charges and the entropy for this black hole and compare the result with the asymptotic structure. As the main purpose, we compute the EPL to testify its universality and then corroborate the thermodynamic laws. In Sec. 3, we will do similar procedure for the BTZ black hole in the MMG theory. Finally, in the last section we discuss about the results and give a brief summary.    
\section{WAdS$_3$ BLACK HOLE IN THE NMG SECTOR}

The NMG is the simplest ghost free, parity-even, forth-derivatives higher curvature gravity in (2+1)-dimensions, which could be inherited to the Gauss-Bonnet gravity counterpart in higher dimensions. Since we can write the Riemann tensor in terms of the Ricci and Ricciscalar tensors in three dimensions. This theory is described by the following action\cite{Bergshoeff:2009hq},
\be\label{NMGA}
I_{NMG}
=\frac{1}{2\kappa }\,\int d^3 x \sqrt{-g}\left(R-2 \Lambda_0+\frac{3}{8m^2}\,R^2-\frac{1}{m^2} R_{\mu\nu}R^{\mu\nu}\right)\,.
\ee
where $\Lambda_0$ is a cosmological constant term of mass dimension 2 and $\kappa=8\pi G_3$ is the Newton's constant. At first sight, it seems that this is a renormalizable theory of gravity based on Ref. \cite{Oda:2009ys}, but shortly after, it has been shown in Ref. \cite{Muneyuki:2012ur} that the theory with these coefficients is not renormalizable compared to unitarity. This theory has some solutions that are not maximally symmetric ($R_{\mu\nu}\neq-\Lambda g_{\mu\nu}$)\cite{Anninos:2008fx,Clement:2009gq},\cite{Clement:2009ka}-\cite{Nam:2010dd}. One of them is the spacelike WAdS$_3$ black hole, which has an asymptotically warped AdS behavior and is not the solution of the equations of motion for pure gravity with a cosmological term.  The line element of the WAdS$_3$ black hole in the ADM formalism\cite{Arnowitt:1962hi} is given in Ref.\cite{Anninos:2008fx} by
 \be \label{wads1} ds^2=- N^2\, l^2\, dt^2+\frac{l^2 dr^2}{4 N^2 K^2}+K^2 \,l^2\,(d\varphi+N_{\varphi} dt)^2\,,\ee
where the functions $N, K$, and $N_{\varphi}$ are
\bea \label{wads2} N^2=\frac{(\nu^2+3)(r-r_{+})(r-r_{-})}{4K^2}\,,\quad N_{\varphi}=\frac{2\nu r-\sqrt{ (\nu^2+3)\,r_{+} r_{-}}}{2 K^2}\,,\nn\\
K^2=\frac{r}{4}\left[3(\nu^2 -1)+(\nu^2+3)(r_{+}+ r_{-})-4\nu \sqrt{ (\nu^2+3)\,r_{+} r_{-}}\right]\,.\eea

The parameter $\nu$ is a constant warped factor, and $r_{+},r_{-}$ are the outer and inner horizons, respectively. Substituting the metric (\ref{wads1}) in the equations of motion for the action (\ref{NMGA}), we can determine the cosmological constant and the warped factor in terms of other constant parameters $m,l$ as
\be\label{LNu} \nu^2=\frac{2\,m^2 l^2+3}{20}\,,\quad \Lambda_0=\frac{4\,m^4 l^4-468\,m^2 l^2+189}{400\,m^4 l^4}\,, \ee
where $m$ is a mass parameter in action (\ref{NMGA}) and $l$ is the length scale in the background (\ref{wads1}). 
\subsection{Conserved charges}
In order to investigate the EPL, we need to obtain the conserved charges of the WAdS$_3$ black hole in this theory. We have found these physical quantities from the canonical first-order formalism, the details of which are given in Ref.\cite{Yekta:2016jhg}(see also Refs. in \cite{Yekta:2016jhg}). The mass and angular momentum of the black hole, which are the conserved charges associated with the Killing vectors $\prt_t$ and $\prt_\phi$, are given by
  \bea \label{wmam}
M\!\!\!&=&\!\!\!\frac{\nu^2 (\nu^2+3)}{G (20\nu^2-3) }\left[r_{+}+r_{-}-\frac{1}{\nu}\sqrt{(\nu^2+3)r_{+}r_{-}}\right],\nn\\ 
J\!\!\!&=&\!\!\!\frac{\nu^3 (\nu^2+3)l}{4G (20\nu^2-3) }\left[\left(r_{+}+r_{-}-\frac{1}{\nu}\sqrt{(\nu^2+3)r_{+}r_{-}}\right)^2-(r_{+}-r_{-})^2\right].
\eea
Here, the ADT formalism\cite{Abbott:1981ff,Deser:2002jk} cannot be used to determine the conserved charges (ADT conserved charges) since this formalism only works for the asymptotically AdS-like solutions in the NMG, such as the BTZ black hole. There is also another approach to compute the conserved charges in the case of NMG model in Ref. \cite{Clement:2009gq}, in which, by intruducing a superangular momentum, we can produce these quantities, as in Refs.\cite{Ghodsi:2010ev}, \cite{Ghodsi:2011ua}, for asymptotically AdS/WAdS solutions.

We can compute the entropy of the WAdS$_3$ black hole by using the Wald's formula\cite{Wald:1993nt},
\bea \label{entw}
S &=& 4\pi A_h\Big(\frac{\delta{\cal L}}
{\delta {\cal R}_{0202}}(g^{00}g^{22})^{-1}\Big)_h\,,
\eea
where $\mathcal{L}$ is the three-dimensional NMG Lagrangian of the action (\ref{NMGA}) and $A_h=2\pi r_{h} $ is the area of the event horizon. So, for the event horizon $r_{h}=r_{+}$, it becomes 
\be \label{ent1} S=\frac{4\pi \nu^2 l}{G(20\nu^2 -3)}\,(2\nu\, r_{+}-\sqrt{(\nu^2 +3)\,r_{+}r_{-}}\,\,)\,.\ee

This result is also consistent with the value that was calculated from the AdS/CFT correspondence\cite{Yekta:2016jhg}. The asymptotic symmetry group of the WAdS$_3$ has the isometry group of the $SL(2,R)\times U(1)$, the first of which is a Virasoro algebra and the latter is a Ka\v{c}-Moody algebra. Using the Sugawara construction\cite{Sugawara:1967rw}, we can find a $SL(2,R)\times SL(2,R)$ symmetry described by two versions of the Virasoro algebras with central charges\cite{Yekta:2016jhg}:
\be \label{cc1}
c_{L}=c_{R}=c=\frac{96 \nu^3 l}{G(\nu^2+3)(20\nu^2-3)}\,.\ee
Therefore, if there were such a two-dimensional CFT at the asymptotic boundary of this solution, we could calculate the entropy from the Cardy's formula\cite{Cardy:1986ie},
\be \label{cardy1} S=\frac{\pi^2 l}{3}(c_{R} T_{R}+c_{L} T_{L})\,,\ee
where $T_{L},T_{R}$ are the left and right temperatures defined in Ref. \cite{Anninos:2008fx}:
\be T_{L}\equiv\frac{\nu^2+3}{8\pi l} (r_{+}-r_{-}),\quad T_{R}\equiv\frac{\nu^2+3}{8\pi l}\left(r_{+}+r_{-}-\frac{1}{\nu}\sqrt{(\nu^2+3)\,r_{+}r_{-}}\right).\ee
However, we can also define the left and right zero mode energies as in Ref. \cite{Ghodsi:2011ua} by
\be \label{ELR} E_{L}=\frac{\pi^2}{6} c_{L} T_{L}^2= \frac{(20\nu^2-3) G}{4\nu l (\nu^2+3)}\,M^2\,,\qquad E_{R}=\frac{\pi^2}{6}\,c_{R} T_{R}^2=\frac{(20\nu^2-3) G}{4\nu l (\nu^2+3)}\,M^2-\frac{J}{l^2}\,,\ee
where, from another Cardy's formula,
\be\label{cardy2} S=2\pi\left(\sqrt{\frac{c_{L} E_{L}}{6}}+\sqrt{\frac{c_{R} E_{R}}{6}}\,\right), \ee
we again arrive at the result in (\ref{ent1}).
\subsection{Thermodynamics}
So far, we have obtained the entropy for the event horizon ($r_{+}$) of the WAdS$_3$ black hole. To fix our convention, we denote its entropy given in (\ref{ent1}) by $S^{+}$. The Killing  vector field which generates the mass is a timelike vector field on the event horizon; thus, the mass has positive value, while the Killing horizon vector field is spacelike inside the black hole event horizon and we assign negative energy or ADM mass ($-M$) to the inner horizon ($r_{-}$), similar to the negative energies within the ergosphere. Classically, the inner Cauchy horizon is perturbatively unstable; the physical implications of this instability have been recently revisited in Ref.\cite{Marolf:2011dj}. Our purpose is to illustrate how the inner horizon could impact a statistical interpretation of the black hole thermodynamics.

The entropy is computed for the spacelike Killing vector field $\xi_{-}=-\frac{4\pi}{\kappa}(\prt_{t}+\Omega \prt_{\phi})$, as
\be\label{ent2} S^{-}=\frac{4\pi \nu^2 l}{G(20\nu^2 -3)}\,(2\nu\, r_{-}-\sqrt{(\nu^2 +3)\,r_{+}r_{-}}\,\,), \ee
which can be calculated similarly from the Cardy-like formula, 
\be \label{cardy3} S^{-}=\frac{\pi^2 l}{3}(c_{R} T_{R}- c_{L} T_{L})\,,\ee
or equivalently by
\be \label{cardy4} S^{-}=2\pi\left(\sqrt{\frac{c_{R} E_{R}}{6}}-\sqrt{\frac{c_{L} E_{L}}{6}}\,\right).\ee

Now we are ready to verify the EPL for the WAdS$_3$ black hole in the NMG sector. From the relations (\ref{ent1}) and (\ref{ent2}), and comparing with conserved charges (\ref{wmam}), we obtain
\be\label{PL} \frac{S^{+} S^{-}}{4\pi^2}=\frac{c}{6}\, J\,,\ee
where, the result is independent of the mass and only depends on the quantized parameter $J$. As a consequence, the EPL is a universal law for this kind of massive gravity theory in the WAdS$_3$ background. Fortunately, the result is also in accordance with that proposed in Ref.\cite{Detournay:2012ug}. It is also consistent with the level matching condition ($E_{L}-E_{R}$) for the modes given in definition (\ref{ELR}).

As a complementary check, we explore the first law of black hole thermodynamics for the inner and outer horizons, separately.
The temperature and angular velocity for the event horizon of the WAdS$_3$ black hole, which is described by (\ref{wads1}) and (\ref{wads2}), are
\be \label{TV1} T^{+}=\frac{\nu^2+3}{4\pi l}\,\frac{r_{+}-r_{-}}{2\nu\, r_{+}-\sqrt{(\nu^2 +3)\,r_{+}r_{-}}}\,,\qquad \Omega^{+}=\frac{2}{l\, (2\nu\, r_{+}-\sqrt{(\nu^2 +3)\,r_{+}r_{-}})}\,,\ee
and for the inner horizon, respectively,
\be \label{TV2} T^{-}=\frac{\nu^2+3}{4\pi l}\,\frac{r_{+}-r_{-}}{2\nu\, r_{-}-\sqrt{(\nu^2 +3)\,r_{+}r_{-}}}\,,\qquad \Omega^{-}=\frac{2}{l\, (2\nu\, r_{-}-\sqrt{(\nu^2 +3)\,r_{+}r_{-}})}\,.\ee

Thus, from the relation (\ref{wmam}) for the mass and angular momentum and the entropies (\ref{ent1}) and (\ref{ent2}), in addition to the above variables, one can see that all of them satisfy the following first laws of the nonextremal black holes thermodynamics:
\be\label{fl} \pm dM = T^{\pm} dS^{\pm}\pm\, \Omega^{\pm} dJ.\ee

We also calculate that the physical parameters are consistent with the Smarr-like formula\cite{Smarr:1972kt} for the WAdS$_3$ black hole\cite{Ghodsi:2011ua},\cite{Yekta:2016jhg}:
\be\label{smarr1} \pm M=T^{\pm} S^{\pm}\pm 2\,\Omega^{\pm} J\,.\ee

\section{BTZ BLACK HOLE IN THE MMG SECTOR}
The other intriguing ghost-free, unitary(tachyon-free), four-derivatives toy model of three-dimensional massive gravity is the MMG theory. Though MMG is a quadratic curvature theory of gravity, the linearization of this theory around an AdS-like solution gives only one massive mode as TMG, so this is called minimal massive gravity\cite{Bergshoeff:2014pca}. The unusual feature of this theory is that it has no Lagrangian in the metric formalism, but is described at the level of the equations of motion by the following equations,
\be \label{MMG} \frac{1}{\mu}\, C_{\mu\nu}+\bar\sigma\, G_{\mu\nu}+\bar\Lambda_{0}\, g_{\mu\nu}=-\frac{\gamma}{\mu^2}\,J_{\mu\nu},\ee  
where $J_{\mu\nu}$ is a symmetric tensor,
\be J_{\mu\nu}=\frac{1}{2 |g|}\, {\varepsilon_{\mu}}^{\rho\lambda}\,{\varepsilon_{\nu}}^{\tau\eta} S_{\rho\tau} S_{\lambda\eta}\,,\ee
and 
\be C_{\mu\nu}=\frac{1}{\sqrt{-|g|}}\, {\varepsilon_{\mu}}^{\tau\eta} D_{\tau} S_{\eta\nu}\,\,,\qquad S_{\mu\nu}=R_{\mu\nu}-\frac14 g_{\mu\nu} R\,, \ee
are the Cotton tensor and Shouten tensor, respectively. 
$\gamma$ is a nonzero dimensionless constant and $\sigma$, $\Lambda_{0}$ in the case of TMG are replaced by $\bar\sigma$, $\bar\Lambda_{0}$ in (\ref{MMG}) for MMG since it is not obvious they should be equal to initial values. It has been shown in Ref.\cite{Yekta:2015gja} that this theory can be described by a 3-form Lagrangian in the Chern-Simons-like formalism as
\be \label{LMMG1} L_{MMG}=-\sigma\,e\cdot R+\,\frac{\Lambda_0}{6}\, e\cdot e\times e+h \cdot T+\frac{1}{2\mu}\left(\omega \cdot d \omega+\frac13\, \omega\cdot \omega\times\omega\right)+\frac{\alpha}{2}\,e\cdot h\times h\,,\ee
where $e$,$h$, and $\omega$ are three 1-form fields and $R,T$ are the 2-form Lorentz curvature and torsion tensors. A well-defined black hole solution for the general three-dimensional gravities, such as MMG in here, is the rotating BTZ black hole. There are also other solutions which can be found in Refs.\cite{Arvanitakis:2015yya,Altas:2015dfa}.

The metric of this black hole in the ADM  formalism is given by 
\be \label{btz1} ds^2=-N^2 dt^2+N^{-2} dr^2+r^2(d\varphi+N_{\varphi} dt)^2\,,\ee
where the functions $N, N_{\varphi}$ are given by 
\be N^2=\frac{(r^2-r_{+}^2)(r^2-r_{-}^2)}{r^2 l^2}\,,\quad N_{\varphi}=\frac{r_{+} r_{-}}{l\, r^2}\,.\ee
Similar to the WAdS$_3$ black hole, $r_{+}$ and $r_{-}$ play the role of the inner and outer horizons of the BTZ black hole, the physical quantities of which can be described in terms of them. Substituting the metric (\ref{btz1}) in Eq. (\ref{MMG}) gives the following constraint on the parameter space of the theory:
\be \bar{\Lambda}_0=-\frac{\gamma+4\,\mu^2 l^2 \bar{\sigma}}{4 \mu^2 l^4}.\ee
\subsection{Conserved charges}
We have computed the conserved charges of this theory for the BTZ black hole in the Hamiltonian formalism\cite{Yekta:2015gja}. Similar calculations from the ADT approach and phase-space formalism have been done in Refs. \cite{Tekin:2014jna},\cite{Setare:2015pva}. The mass and angular momentum of the black hole are computed as
\bea \label{eambtz}
M\!\!\!&=&\!\!\!\frac{1}{4G}\left[(\sigma+\alpha C) \frac{r_{+}^2+r_{-}^2}{2 l^2}+\frac{1}{\mu l}\,\frac{r_{+} r_{-}}{l^2}\right]\,,\nn\\ 
J\!\!\!&=&\!\!\!\frac{1}{4G}\left[(\sigma+\alpha C)\frac{r_{+} r_{-}}{l} +\frac{1}{\mu l}\,\frac{r_{+}^2+r_{-}^2}{2 l}\right]\,,
\eea
where the constants $\sigma$, $\alpha$, and $C$ are introduced in the relations (2.12) and (4.7) of Ref.\cite{Yekta:2015gja}. Since this theory has no Lagrangian in metric formalism, we cannot utilize the Wald's formula for calculating the entropy. On the other hand, there is a phase-space formalism which gives the correct results\cite{Solodukhin:2005ah},\cite{Tachikawa:2006sz}. The details of the calculation are given in Ref.\cite{Setare:2015pva}. 
Therefore, the entropy of the BTZ black hole in the MMG theory is
\be \label{ent3} S^{+}=\frac{A_{h}}{4G}\left[(\sigma+\alpha C)+\frac{1}{\mu l} \frac{r_{-}}{r_{+}}\right]\,,\ee
where $A_{h}=2\pi r_{+}$ is the area of outer event horizon. The BTZ black holes are asymptotically AdS$_3$ solutions, so from the AdS/CFT conjecture\cite{Maldacena:1997re,Witten:1998qj} and the Brown-Henneaux approach\cite{Brown:1986nw}, we can find the central charges of the algebras determining the asymptotic symmetry group. There are two versions of the Virasoro algebra with the left and right central charges:
\be \label{ccbtz} c_{L}=\frac{3l}{2G}\left(\sigma+\alpha C-\frac{1}{\mu l}\right)\,,\qquad c_{R}=\frac{3l}{2G}\left(\sigma+\alpha C+\frac{1}{\mu l}\right)\,.\ee
In the CFT side, by defining the left and right temperatures for the BTZ black hole \cite{Maldacena:1998bw} by
\be T_{L}\equiv\frac{r_{+}-r_{-}}{2\pi l^2}\quad\,,\quad T_{R}\equiv\frac{r_{+}+r_{-}}{2\pi l^2}\,,\ee
and the Cardy's formula (\ref{cardy1}) and the values in (\ref{ccbtz}), we again obtain the entropy (\ref{ent3}). The zero-mode energies of the left and right sectors in the asymptotic theory are also defined as the following
\be\label{ELRbtz} E_{L}=\frac{\pi^2}{6} c_{L} T_{L}^2=\frac{M\,l-J}{2\,l} \,,\qquad E_{R}=\frac{\pi^2}{6}\,c_{R} T_{R}^2=\frac{M\,l+J}{2\,l}\,.\ee 
 
\subsection{Thermodynamics}
The thermodynamic properties, including the temperature (surface gravity) and angular velocity at each horizon, can be expected to play a central role in understanding the entropy at the microscopic level. Therefore, for the BTZ black hole, the temperature and angular velocity of the event horizon $r_{+}$ are given by
\be \label{TVbtz1} T^{+}=\frac{1}{2\pi}\,\kappa\,|_{r=r_{+}}=\frac{r_{+}}{2\pi l^2}\,(1-\frac{r_{-}^2}{r_{+}^2})\,,\qquad \Omega^{+}=\frac{1}{l}\,N_{\varphi}|_{r=r_{+}}=\frac{r_{-}}{l\, r_{+}}\,.\ee
where $\kappa$ is the surface gravity on the horizon. We need the values of the parameters for the inner Cauchy horizon to verify the universality of the EPL. The energy(mass) becomes negative inside the event horizon and the entropy for the spacelike Killing vector field becomes
\be \label{ent4} S^{-}=\frac{A_{h}}{4G}\left[(\sigma+\alpha C)+\frac{1}{\mu l} \frac{r_{+}}{r_{-}}\right]\,,\ee
where $A_{h}=2\pi r_{-}$ is the area of the inner horizon. So the product of the entropies (\ref{ent3}) and (\ref{ent4}) becomes
\be \label{eplbtz} \frac{S^{+} S^{-}}{4\pi^2}=\frac{M l+\mu\,l (\sigma+\alpha C)\,J}{8\mu G}\,. \ee

This result shows the failure of the universality of the EPL. That is, the product is not independent of the mass of the black hole, as it must be, but it can be seen that, in the limit $\mu\rightarrow \infty$, the result only depends on the angular momentum $J$. Thus, we expect when there is a holographic gravitational anomaly\cite{Kraus:2005zm} (such as TMG), this rule has been broken. Similar to what happens for the central function, $ c=\frac{l}{2G}\,g_{\mu\nu}\,\frac{\prt \mathcal L_3}{\prt R_{\mu\nu}}$, in the holographic calculations\cite{Kraus:2005vz},
\be \label{hga} c_{L}-c_{R}=-\frac{3}{\mu\,G}\,.\ee
We can rewrite the EPL (\ref{eplbtz}) in terms of the energies in (\ref{ELRbtz}) with
\be\label{eplbtz2} \frac{S^{+} S^{-}}{4\pi^2}=\frac{l}{6}\,(c_{R} E_{R}-c_{L} E_{L})\,,\ee 
or, analogously, with
\be \label{eplbtz3} \frac{S^{+} S^{-}}{4\pi^2}=\frac{l}{12}\,\Big( (c_{R}+c_{L})\,( E_{R}-E_{L})+(c_{R}-c_{L})\,( E_{R}+E_{L})\Big)\,,\ee
where again, in the limit $\mu\rightarrow\infty$ from (\ref{hga}), the second term vanishes, the central charges in (\ref{ccbtz}) become equal, and the EPL (\ref{eplbtz3}) leads to the relation (\ref{eplbtz2}). 
In the final stage, we consider the first law of thermodynamics for the BTZ black hole in MMG. We need to compute the temperature and angular velocity of the inner horizon, which become
\be \label{TVbtz2} T^{-}=\frac{r_{+}^2-r_{-}^2}{2\pi l^2 r_{-}}\,,\qquad \Omega^{-}=\frac{r_{+}}{l\, r_{-}}\,.\ee
Now from the mass and angular momentum (\ref{eambtz}) and the entropies (\ref{ent3}) and (\ref{ent4}) together the intensive quantities computed at the outer (\ref{TVbtz1}) and inner (\ref{TVbtz2}) horizons, the first law holds like the relation (\ref{fl}). Another thermodynamical calculation for the general spinning BTZ black holes has been done in Ref.\cite{Pradhan:2015ela}. The Smarr-like relations also hold for the BTZ black hole, in the form
\be\label{smarr2} \pm M=\frac12 \,T^{\pm} S^{\pm}\pm \,\Omega^{\pm} J\,.\ee

\section{DISCUSSION}
In this paper, we have investigated the universality of the product law of  inner Cauchy and outer event horizon's entropies for two particular branches of the nonextremal black hole solutions in the three-dimensional gravity toy models. In the first case, we studied the rotating spacelike warped AdS$_3$ black hole in the quadratic curvature gravity, which is called new massive gravity, and in the second case, we considered the rotating BTZ black hole in the minimal massive gravity as a minimal extension of the topologically massive gravity, although it is free of ghost and tachyon. In both inspections, we have used the canonical Hamiltonian first-order formalism to calculate the conserved quantities of the black holes. Specially, we compute the mass and angular momentum, which are important in searching the EPL for the solutions. 

On the thermodynamic side, not only did we obtain the quantities for the event outer horizon, but we also computed them for the inner Cauchy horizon, where the ADM energy or mass becomes negative. We have shown that for the WAdS$_3$ in the NMG sector, the universality holds and the product is independent of the mass, whereas for the BTZ in the MMG theory, it fails. We have seen that this violation comes from the gravitational anomalous term with the coupling constant $\frac{1}{\mu}$\,, such that in the limit $\mu\rightarrow\infty$, the product becomes mass independent and holds the level matching condition ($E_{R}-E_{L}$). Having used the asymptotic symmetry group analysis, we have checked that the entropies satisfy the Cardy formula for the CFT dual theories. Finally, we show that the thermodynamical quantities of the inner and outer horizons satisfy the first law of black hole thermodynamics and the Smarr formula for the mass of the black holes.    


\end{document}